\documentstyle[aaspp4]{article}

\begin{document}

\title{Statistical properties of SGR 1900+14 bursts}

\author{Ersin {G\"o\u{g}\"u\c{s}}\altaffilmark{1,3}, 
Peter M. Woods\altaffilmark{1,3}, Chryssa Kouveliotou\altaffilmark{2,3},
Jan van Paradijs\altaffilmark{1,4}, Michael S. Briggs\altaffilmark{1,3},
Robert C. Duncan\altaffilmark{5}, Christopher Thompson\altaffilmark{6}}

\altaffiltext{1}{Department of Physics, University of Alabama in Huntsville, 
       Huntsville, AL 35899} 
\altaffiltext{2}{Universities Space Research Association}
\altaffiltext{3}{NASA Marshall Space Flight Center, SD-50, Huntsville, AL 35812}
\altaffiltext{4}{Astronomical Institute "Anton Pannekoek", University of
      Amsterdam, 403 Kruislaan, 1098 SJ Amsterdam, NL} 
\altaffiltext{5}{Department of Astronomy, University of Texas, RLM 15.308, 
      Austin, TX 78712-1083}
\altaffiltext{6}{Department of Physics and Astronomy, University of North
      Carolina, Philips Hall, Chapel Hill, NC, 27599-3255}

\authoremail{Ersin.Gogus@msfc.nasa.gov}

\begin{abstract}

We study the statistics of soft gamma repeater (SGR) bursts, using
a data base of 187 events detected with BATSE and 837 events detected with 
RXTE PCA, all from SGR 1900+14 during its 1998-1999 active phase. 
We find that the fluence or energy distribution of bursts is consistent 
with a power law of index 1.66, over 4 orders of magnitude.
This scale-free distribution resembles the Gutenberg-Richter Law for
earthquakes, and gives evidence for self-organized criticality in SGRs. 
The distribution of time intervals between successive bursts from 
SGR 1900+14 is consistent with a log-normal distribution.  There is no 
correlation between burst intensity 
and the waiting times till the next burst, but there is some
evidence for a correlation between burst intensity and the time elapsed
since the previous burst.  We also find a correlation between the
duration and the energy of the bursts, but with significant scatter. 
In all these statistical properties, SGR bursts resemble earthquakes 
and solar flares more closely than they resemble any
known accretion-powered or nuclear-powered phenomena. Thus our analysis lends
support to the hypothesis that the energy source for SGR bursts is internal
to the neutron star, and plausibly magnetic.

\end{abstract}


\section{Introduction}

At least three of the four currently-known soft gamma repeaters are 
associated with slowly 
rotating, extremely magnetized neutron stars located within young 
supernova remnants (Kouveliotou et al. 1998, 1999). They are characterized 
by the recurrent emission of gamma-ray bursts with relatively soft spectra 
(resembling optically-thin
thermal bremsstrahlung at $kT \sim 20$--40 keV) and short durations
($\sim$ 0.1 s) (Kouveliotou 1995). Thompson and Duncan (1995) suggested that
these bursts are due to neutron star crust fractures, driven by the 
stress of an evolving, ultra-strong magnetic field, $B \gtrsim 10^{14}$
Gauss.  

Cheng, Epstein, Guyer \& Young (1996) \markcite{cheng96} observed that 
particular statistical 
properties of a sample of 111 SGR events from SGR 1806-20 are quite similar
to those of earthquakes (EQ). These properties include the distribution of 
event energies, which follow a power law 
dN $\propto$ E$^{-\gamma}$~dE 
with an exponent, $\gamma$ = 1.6. A similar distribution was 
obtained empirically by Gutenberg and Richter (1956a\markcite{gr56a}; 
1965\markcite{gr65})
for the distribution of EQ energies, with power law index 
$\gamma_{EQ}$=$1.6 \pm 0.2$;
and in computer simulations of fractures in a stressed, elastic medium
(Katz 1986\markcite{katz86}).
The distribution of time intervals between successive SGR 1806-20
events is well described by a log-normal distribution analogous to the
waiting times distribution of microglitches seen in the Vela pulsar 
(see Hurley et al.~1994)\markcite{hur94}. 
Cheng et al.~(1996) also showed that cumulative 
waiting time distributions of SGR 1806-20 and EQ events are similar.
These results support the idea that SGR bursts are caused by starquakes,
as expected to occur in the crusts of magnetically-powered neutron stars, 
or ``magnetars" (Duncan \& Thompson 1992\markcite{dt92}; Thompson and Duncan
1995\markcite{td95}, 1996\markcite{td96}).

In May 1998, SGR 1900+14 became extremely active after a long period during
which only sporadic activity occurred (Kouveliotou 1993\markcite{kou93}).
In the period from May
1998 until January 1999 a total of 200 events were detected (Woods et al.
1999b)
with the Burst and Transient Source Experiment (BATSE) aboard the Compton
Gamma Ray Observatory (CGRO). Out of these 200 events, 63 led to an
on-board trigger. 
The sudden change in source activity initiated a series of 
Rossi X-ray Timing Explorer (RXTE)
observations between May 31 and December 21, 1998. During these
observations, 837 bursts from SGR 1900+14 were detected with the
Proportional Counter Array (PCA). As noted by Kouveliotou et al.~(1998) for
SGR 1806-20, the bursts occur in an apparently irregular temporal pattern. This
is also true for SGR 1900+14 bursts\setcounter{footnote}{0}
\footnote{Some examples of irregular 
temporal pattern of SGR 1900+14 bursts can be
seen at \tt{http://gammaray.msfc.nasa.gov/batse/sgr/sgr1900/}}. 

In this {\it Letter}, we study the statistics of SGR bursting 
using the new measurements of SGR 1900+14.  \ Our data base of events 
is larger by $\sim 10$ than that of previous statistical studies,
and extends over a larger dynamic range in burst energy (or fluence) 
by $\sim 10^2$. 

\section{BATSE Observations} 

The BATSE instrument is made up of 8 identical detector modules located on
each
corner of the CGRO. Each module contains a large
area detector (LAD) and a spectroscopy detector (SD). In our analysis, we
have
used DISCriminator LAD (DISCLA) data with coarse energy resolution (4
channels
covering E$>$25 keV), Spectroscopy Time-Tagged Event (STTE) data and
Spectroscopy High Energy Resolution Burst (SHERB) data with fine energy
binning
(256 channels).
A detailed description of BATSE instrumentation
and data types can be found in Fishman et al.~(1989)\markcite{fish89}.

BATSE was triggered by SGR 1900+14 bursts 63 times between May 1998 and
January
1999. For 22 of the brightest  events, we obtained STTE or SHERB data with 
detailed spectral information. We fit the
background subtracted source spectra to optically-thin thermal
bremsstrahlung
(OTTB) and power law models. The OTTB model, F(E)$\propto$ 
E$^{-1}$$\exp$($-$E/kT), provides suitable fits 
(0.83 $<$ $\chi^{2}_\nu$ $<$ 1.32) to  all of the 
event spectra with temperatures ranging between 21.0 and 46.9 keV. The
power law model failed to fit most of the spectra. The mean of the
OTTB temperatures for this sample of 22 events, appropriately weighted by  
uncertainties, is $25.7 \pm 0.2$ keV.

Woods et al.~(1999b) performed an extensive search for untriggered BATSE
events
from SGR 1900+14. They found, in addition to the 63 triggered events, 
137 untriggered burst events between 24 May 1998 and 3 February 1999. In
this study we selected the 187 BATSE events (triggered and untriggered)
which 
had
DISCLA data. This data type is read out continuously (with the exception of
data
telemetry gaps) and therefore is available for the largest sample of events.
We have excluded events which occurred in data gaps, events that were too
weak
to fit, and four events due to their distinction from typical SGR activity.
The events on 1998 October 22 and 1999 January 10 
with relatively hard spectra (Woods et al.~1999c), and the multi-episodic 
events on 1998 May 30 and September 1 ({G\"o\u{g}\"u\c{s}} et al.~1999b)
will 
be discussed elsewhere.
Given the long DISCLA data integration time (1.024 s)
relative to typical burst durations ($\sim$ 0.1 s), we could only estimate
the
fluence for each event. In order to determine the fluence of
each burst, we fit the background-subtracted source spectrum to the 
OTTB model with a fixed kT of 25.7 keV, a reasonable choice considering the 
fairly narrow kT distribution of the triggered bursts.
We find that the fluences of SGR 1900+14 bursts observed with BATSE 
range between $2 \times 10^{-8}$ and
$2.5 \times 10^{-5}$ ergs cm$^{-2}$. For an estimated distance to SGR
1900+14 
of 7 kpc (Vasisht et al.~1994)\markcite{vas94}, and assuming
isotropic emission, the corresponding  
energy range is $1.1 \times 10^{38}$ -- $1.5 \times 10^{41}$ ergs.

\section{PCA Observations}

RXTE observed SGR 1900+14 for a total exposure time  of $\sim$ 180 ks
between 
May 1998 and December 1998. In this
work, we have analyzed data from 32 pointed observations with the
PCA. We performed an automated burst search similar to the one
used on BATSE data described by Woods et al.~(1999b). Using Standard 1 data 
(2-60keV) 
for all times where the source was above the Earth's horizon by more than 5
${^\circ}$, we searched for bursts using the following methodology. For each
0.125 s bin, a background count rate was estimated by fitting a first order
polynomial
to 5 s of data before and after each bin with a 3 s gap between the bin
searched
and the background intervals. Bins with count rates exceeding 1000 counts
s$^{-1}$ were assumed to contain burst emission and were excluded from
background intervals. At the beginning(end) of each continuous stretch of
data,
extrapolations of background fits after(before) the bin were used to
estimate
the background count rate within the bin searched. A burst was defined as
any
continuous set of bins with count rates in excess of 5.5 $\sigma$ above the
estimated background.  
The count fluence of each burst was measured by simply integrating the
background-subtracted counts over the bins covering the event.

In order to compare integrated counts obtained with the PCA and
BATSE fluences, we determined a conversion factor between each PCA count and
BATSE fluence. We assume a constant spectral model (OTTB with kT=25.7 keV).
First, we searched for simultaneous bursts observed with both
instruments and we found 13 events ( 6 triggered events, 3 in the read-out
of triggered events and 4 untriggered events in BATSE). We then computed the
ratio of the BATSE fluence of each simultaneous event to the PCA counts,
which ranges  over a factor of $\sim$ 2 between  
$4.65 \times 10^{-12}$ and $1.15 \times 10^{-11}$ ergs cm$^{-2}$counts$^{-1}$.
The weighted mean of the ratios for SGR 1900+14 is  
$5.45 \times 10^{-12}$ ergs cm$^{-2}$counts$^{-1}$ and the standard
deviation, $\sigma$ = $2 \times 10^{-12}$ ergs cm$^{-2}$counts$^{-1}$. 
Invoking this conversion factor, the
fluence of the bursts from PCA extends from $1.2 \times 10^{-10}$ to 
$3.3 \times 10^{-7}$ ergs cm$^{-2}$ (in the BATSE energy range, E $>$ 25
keV) and the burst energies range from $7 \times 10^{35}$ to 
$2 \times 10^{39}$ ergs.

\section{Statistical Data Analysis}

{\it{i) Fluence distributions}}: The fluences of BATSE
bursts were binned in equally spaced logarithmic fluence steps ( $dN/d\log E$) 
(Fig.1). Using a standard least squares fitting method, 
we fit a power law model to data between
$5.0 \times10^{-8}$ and $2.5 \times 10^{-6}$ ergs cm$^{-2}$.
Bursts at the low end of the distribution were excluded because of
diminished
detection efficiency as well as at the high end due to undersampling of the
intrinsic distribution.
The power law exponent obtained is $0.65 \pm 0.08$ (solid line passing
through BATSE data in Fig.1), which corresponds to dN $\propto$ 
E$^{-1.65}$~dE. We also employed a maximum likelihood analysis, instead of the 
least squares method, to fit
a power law model to the unbinned fluence values within the same interval of
fluences. This method yields $\gamma$=$1.66 \pm _{0.12}^{0.13}$ 
for the energy exponent which agrees well with the least squares fit.

Using the conversion factor we derived from RXTE counts to BATSE fluence 
for SGR 1900+14, we determined the fluence of each RXTE burst (in the BATSE
energy range) and distributed
them over the same logarithmic fluence steps (Fig.1). We first fit binned
RXTE fluences between $1.6 \times 10^{-10}$ and $3.3 \times 10^{-7}$ ergs
cm$^{-2}$ to a power law model using least squares method which gives an 
exponent value of $0.64 \pm 0.04$ (solid line passing through RXTE data in 
Fig.1). The unbinned fluences were then fit to the same model using the maximum 
likelihood method obtaining $1.66 \pm 0.05$ for the power law exponent.
Combined
RXTE and BATSE fluences range from $1.2 \times 10^{-10}$ to $2.5 \times 
10^{-5}$ ergs cm$^{-2}$ (Fig.1) which demonstrates that power law
distribution of energies with an exponent $\gamma \approx 1.66$ is valid for 
SGR 1900+14 over 4 orders of magnitude.

{\it{ii) Waiting times statistics}}: 
We have measured the waiting times ($\Delta$T)
between successive bursts, uninterupted by Earth occultation and data gaps,
for 779 events. Fig 2 shows the distribution of waiting times which range from
0.25 to 1421 s.
We fit the ($\Delta$T)-distribution to a log-normal function and found
a peak at $\sim$ 49 s. The solid line in Fig.2 shows the interval used to
fit and the dashed lines are the extrapolations of log-normal distribution. We
do not include waiting times less than 2 s since these bursts appear to be
double peaked events in which the second burst peak appears shortly after the 
first one, although recorded as two distinct bursts. 
We were unable to generate a $\Delta$T-distribution for BATSE bursts due to
the much smaller number of events which occurred during a single orbital window.

In order to investigate any relations between waiting times till the next
burst ($\Delta$$T^{+}$) and the intensity of the bursts, we divided the 779 
events sample into 8 intensity intervals each of which contains approximately 
100 events. We fit the $\Delta$$T^{+}$-distribution 
to a log-normal distribution and determined the
mean-{$\Delta$$T^{+}$} (i.e. where the fitted log-normal distribution
peaks), 
and 
the mean counts for each of the 8 groups. We show in Fig.3-a that there is
no
correlation between $\Delta$$T^{+}$ and energy of the bursts (Spearman
rank-order correlation coefficient, $\rho$ = 0.05 and the probability that
this
correlation occurs by a random data set, P = 0.91). 
We also searched for the relation between the elapsed times since the
previous
burst ($\Delta$$T^{-}$) and the intensity of the bursts. Similar to the
previous
case, we sub-divided the events into 8 intensity intervals and determined 
mean-{$\Delta$$T^{-}$} by fitting to a log-normal distribution and mean
counts 
for each group individually.
Fig.3-b shows that there appears to be an anti-correlation between 
mean-{$\Delta$$T^{-}$} and burst energy ($\rho$ = $-0.93$, P = 8$\times 
10^{-4}$). 

{\it{iii) Burst durations}}:
Gutenberg and Richter (1956a; 1956b) demonstrated 
that there is a power law relation 
between the magnitude, or energy of the EQ events
and the durations of the strong
motion at short distances from an EQ region. In order to investigate if a
similar correlation exists for SGR events, we selected all 679 PCA bursts
from 
the most active period of SGR 1900+14. In order to
determine the durations of the bursts accurately, we used event mode PCA
data 
with 1 ms time resolution. For 281 of the bursts selected, we obtained
t$_{90}$
durations (Koshut et al.~1996) of the bursts. Fig.4 shows that burst
energies 
and durations are correlated ($\rho$=0.54, P$\sim$$10^{-24}$), although
there 
is a significant spread of fluences at a given duration.

\section{Discussion}

The power-law size distribution of SGR 1900+14 bursts with an 
index $\gamma$ = 1.66  is
similar to those found for SGR 1806-20 (Cheng et al.~1996) and SGR 1627-41
(Woods et al.~1999a). The lack of a high energy cut-off in the differential 
size distribution indicates that the highest energy events are not well
sampled in our distribution.

The distribution of waiting times between successive SGR 1900+14 bursts is
characterized by a log-normal function, similar to that of SGR 1806-20
(Hurley et al.~1994). Waiting times between SGR 1900+14 bursts are on average
shorter than those
of SGR 1806-20 since all SGR 1900+14 bursts occurred during the most active
period of the source.
There is no correlation between the intensity of the burst and the waiting
time until the following burst. This result agrees well with the results of 
Laros et al.~(1987) for SGR 1806-20 and distinguishes the physical mechanism
of 
SGR 1900+14 bursts from that of type II X-ray bursts from the Rapid Burster 
(Lewin et al.~1976) in which the burst energy is proportional to the waiting
time till the next burst. 
We find an anti-correlation between the intensity of the bursts and the
waiting time since the previous bursts. This is very different from type I
X-ray bursts (thermonuclear flashes, see Lewin, Van Paradijs and Taam 1993) for
which there is a rough positive correlation.
  
There is evidence of a positive correlation between the energy and the 
duration of SGR 1900+14 bursts. Similar behavior was also observed 
for EQs (Gutenberg \& Richter 1956a, 1956b) and solar flares 
(Lu et al.~1993\markcite{lu93}).

The EQ size distribution appears to be a power law with an exponent between
1.4-1.8 independent of geographic location (Gutenberg \& Richter
1956a,1965; 
Lay \& Wallace 1995). Using data taken from the Solar Maximum Mission (SMM)
Crosby et al.~(1993) found a power law size distribution for 12000 solar
flares 
with exponents ranging between 1.53 and 1.73. 
The SMM results have been confirmed by the results from International
Cometary
Explorer (ICE) for 4350 flares that finds an exponent of 1.6 (Lu et
al.~1993).
Gershberg and Shakhovskaya (1983) found that the size distribution of
stellar
flares from 23 stars display power law with exponent between 1.5 and 2.1.

Chen et al.~(1991)\markcite{chen91} argued that EQ 
dynamics is described by a self-organized critical system.
Crosby et al.~(1993) similarly suggested that the size distribution of solar
flares reflects an underlying system in a state of self-organized
criticality 
(see Bak et al.~1988\markcite{bak88}) which states that many composite
systems will self-organize to a critical state in which a small perturbation
can trigger a chain reaction that affects any number of elements within the
system.

We have been unable to find clear results in the literature on the
distribution of waiting times between
successive solar flares or EQs. Wheatland et al.~(1998)\markcite{wheat98} 
predicted that the distribution of waiting times of solar flares 
displays a power law, while Biesecker (1994)\markcite{bies94} proposed that
it is consistent with a time-dependent Poisson process. Nishenko and Bulland
(1987) showed that waiting time distribution of large EQs is well described by 
a log-normal function. In recent work by Nadeau and McEvilly (1999) there is
an evidence of log-normal distribution of waiting times between micro EQs. 
 
The large number of bursts in our samples allow us to stringently test
the power law size distribution proposed by
Cheng et al.~(1996); we find that the size distribution of SGR 1900+14
bursts follows a power-law of index 1.66 over more than four orders of
magnitude in burst fluence.
This behavior, along with a log-normal waiting time distribution and 
energy-correlated burst
durations, are characteristics of self-organized critical systems in
general, and earthquakes and solar flares in particular.
In the magnetar model, the triggering mechanism for SGR bursts is a hybrid
of starquakes and magnetically-powered flares (Thompson \& Duncan 1995).  
When magnetic stresses induce elastic strains in the crust, the 
stored potential energy is predominantly magnetic rather than elastic.
In contrast with an EQ, this allows much of the energy to be released
directly into a propagating disturbance of the external magnetic 
field\footnote{Small scale fractures occuring deep in
the crust excite internal seismic waves that couple only indirectly
to external Alfv\'en modes.} of the neutron star;  and in contrast 
with a solar flare it is the rigidity of the crust that provides a gate 
or trigger for the energy release.  The extended power-law distribution
of burst fluences suggests that the average radiative efficiency does not
vary significantly over four orders of magnitude in burst energy, and
provides a strong constraint on burst emission models (Thompson et al.~1999). 

\acknowledgments

We thank Dr. Robert D. Preece for adjusting WINGSPAN software to
process untriggered BATSE events and Dr. Markus J. Aschwanden for helpful
discussions on solar flares. We acknowledge support from 
the cooperative agreement NCC 8-65 (EG); 
NASA grants NAG5-3674 and NAG5-7060 (JvP); Texas Advanced
Research Project grant ARP-028 and NASA grant NAG5-8381 (RCD).

\newpage

\begin{figure}[h]
\plotone{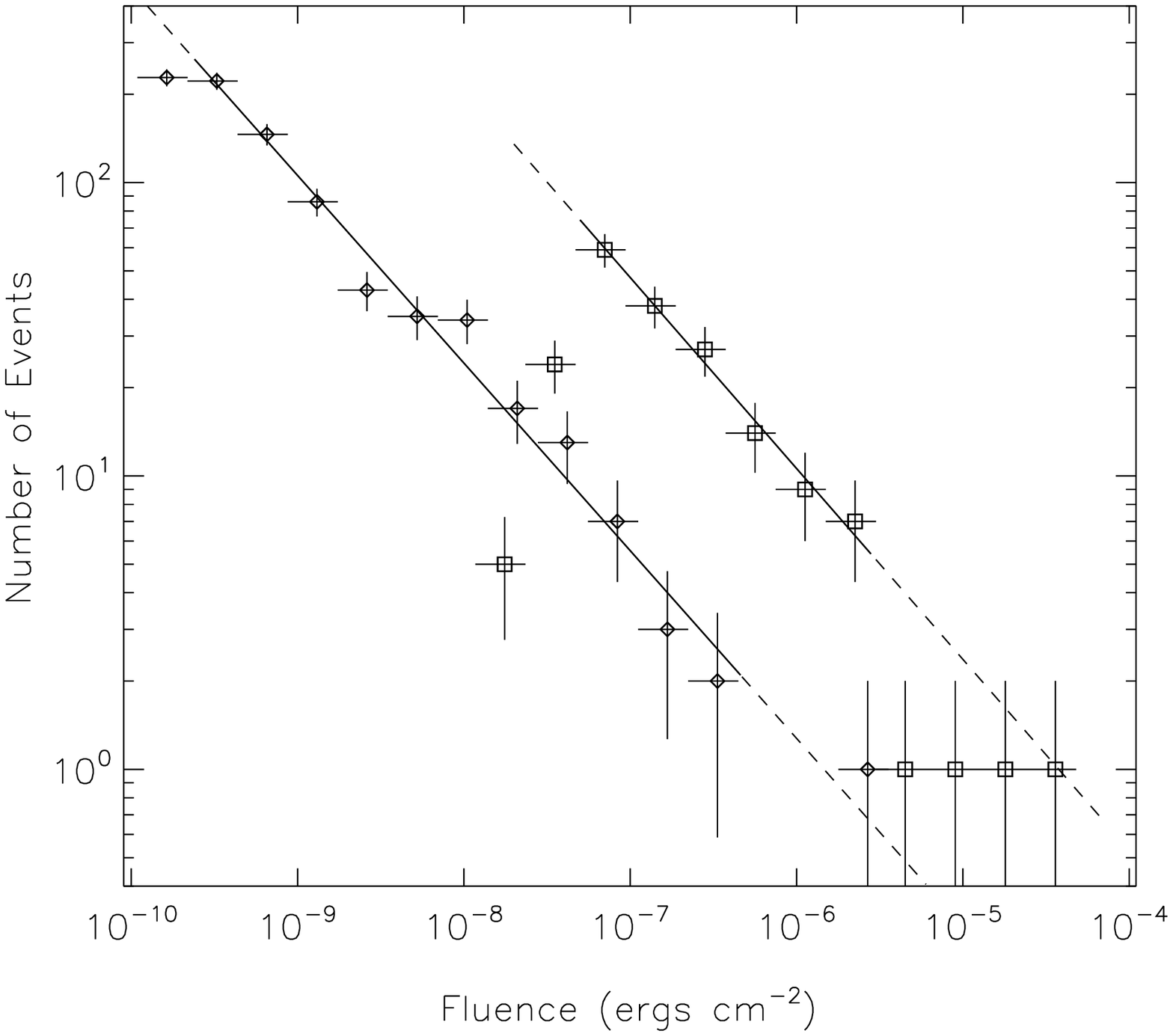}
\caption{Differential distribution of the fluences of bursts from SGR
1900+14 as
measured with RXTE (diamonds) and BATSE 
(squares). The solid lines denote the interval where used in the fit 
and the dashed lines are the extrapolations of the model.} \label{onebarrel}
\end{figure}

\newpage

\begin{figure}[h]
\plotone{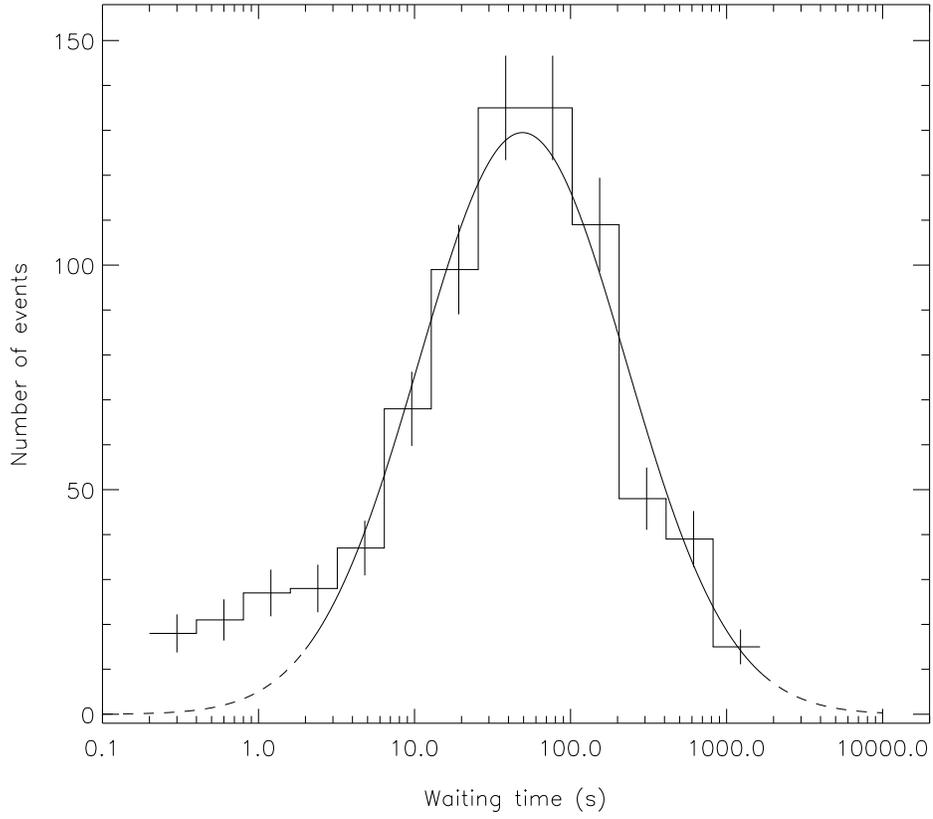}
\caption{Distribution of the waiting times between successive RXTE PCA 
bursts from SGR 1900+14. The line shows the best fit log-normal funtion. The
solid portion of the line indicates the data used in the fit. The excess of
short intervals above the model is due to the double peaked events explained
in
the text.}
\label{onebarrel}
\end{figure}

\newpage

\begin{figure}[h]
\plotone{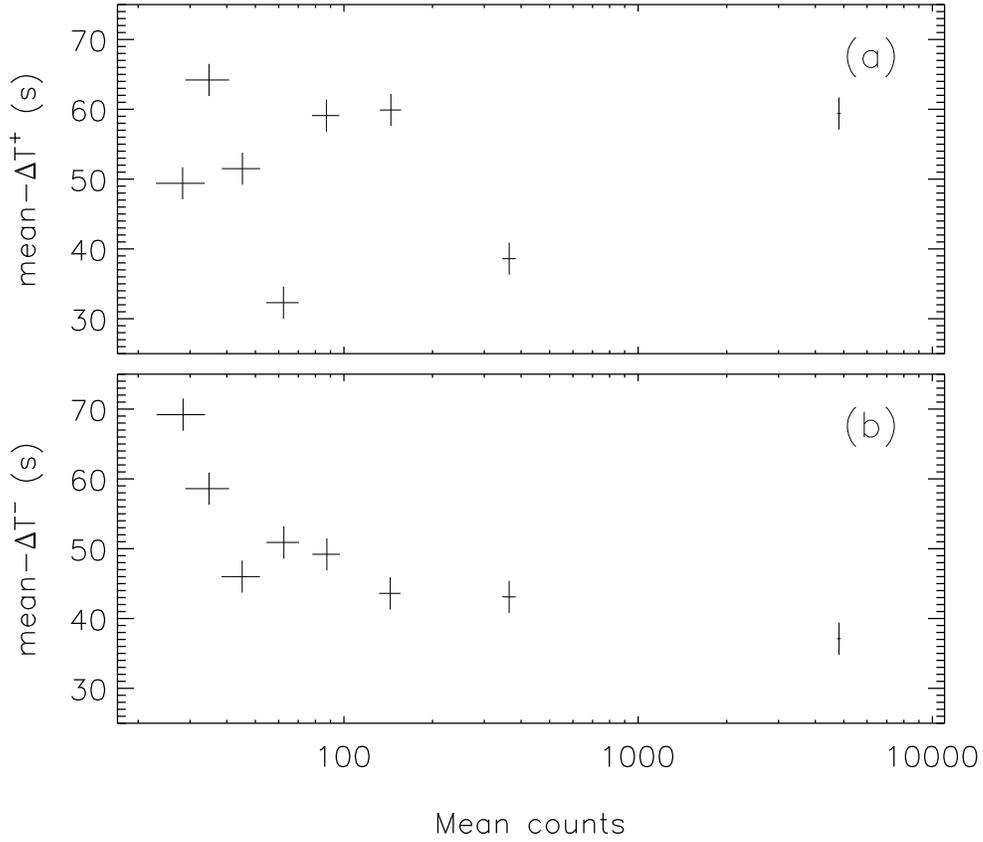}
\caption{Plots of mean waiting times till the next burst ($\Delta$$T^{+}$) 
vs mean
counts (a) which does not show any correlation ($\rho$ = 0.05) and mean
elapsed 
times since the previous burst ($\Delta$$T^{-}$) vs mean counts (b) which
shows
a strong anti-correlation ($\rho$ = $-0.93$).} \label{onebarrel}
\end{figure}

\newpage

\begin{figure}[h]
\plotone{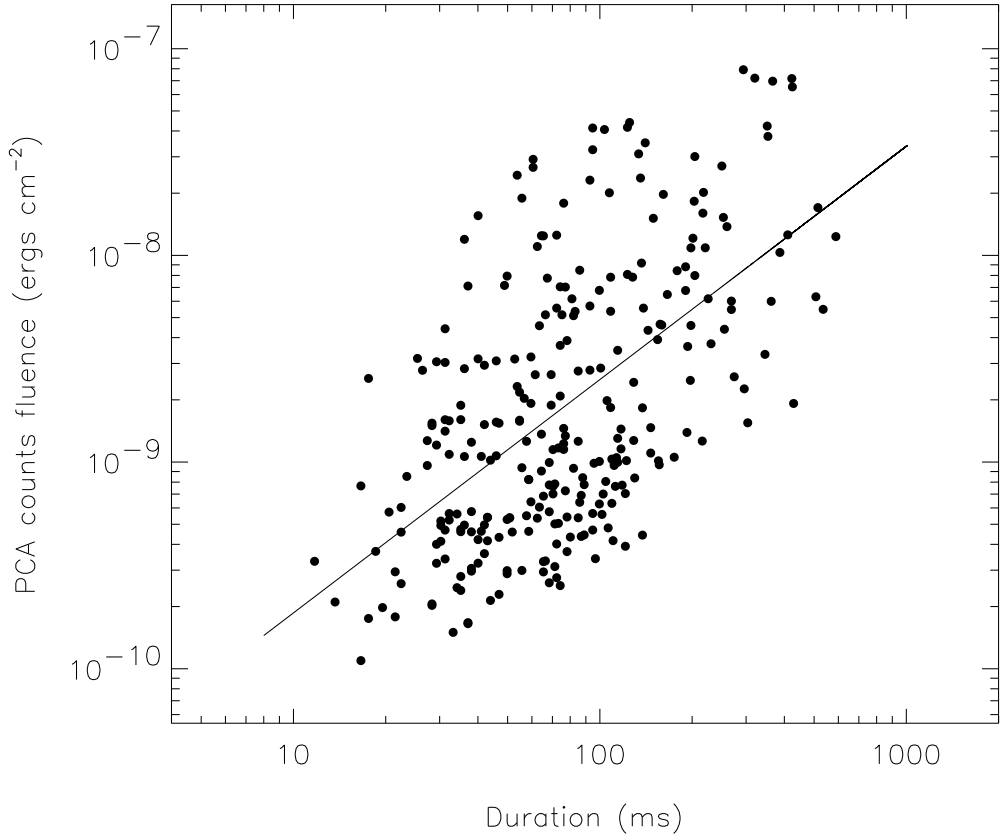}
\caption{Scatter plot of the PCA fluence vs duration for 281 SGR 1900+14 
bursts which shows a correlation between them ($\rho$ = 0.54). The solid 
line is a power law with an exponent 1.13 
obtained using via least squares fitting.} \label{onebarrel}.
\end{figure}

\end{document}